\documentclass{article}
\usepackage{amssymb,bm}
\usepackage{amsmath}
\usepackage{amsthm}
\usepackage{amsfonts}
\usepackage{mathrsfs}
\usepackage{appendix}
\usepackage{hyperref}
\usepackage{authblk}
\usepackage{color}
\usepackage{optidef}

\newtheorem{defn}{Definition}[section]

\theoremstyle{remark}

\usepackage{graphics}
\usepackage{graphicx}
\usepackage{geometry}
\usepackage{algorithm}
\usepackage{algorithmic}
\hypersetup{colorlinks,linkcolor={blue},citecolor={blue},urlcolor={red}}

\usepackage[english]{babel}

\usepackage{cite}
\usepackage{threeparttable}
\usepackage{float}
\floatstyle{plaintop}
\restylefloat{table}

\floatstyle{plain}
\restylefloat{figure}
\usepackage{placeins}

\begin{document}

\title{To AI or not to AI, to Buy Local or not to Buy Local: A Mathematical Theory of Real Price {\footnote{Weiyu Xu co-designed this research, and performed mathematical modeling and derivations. Huan Cai co-designed this research and provided economic interpretations. Catherine Xu derived the mathematical equations presented in Section \ref{subsection_generaln}, and wrote down all the mathematical equations in Section \ref{realprice_dynmaic} in LaTex.
}}

\author{Huan Cai {\footnote { Cornell College, Mount Vernon, IA, 52314.}}\,~~
Catherine Xu {\footnote{
Iowa City Math Circle and Club, Iowa.}}\,~~
 Weiyu Xu {\footnote{
Department of Electrical and Computer Engineering, University of Iowa, Iowa City, IA 52242.  Email: \texttt{weiyu-xu@uiowa.edu}}}
}}

\maketitle

\begin{abstract}
In the past several decades, the world's economy has become increasingly globalized. On the other hand, there are also ideas advocating the practice of ``buy local'', by which people buy locally produced goods and services rather than those produced farther away. 
In this paper, we establish a mathematical theory of real price that determines the optimal global versus local spending of an agent which achieves the agent's optimal tradeoff between spending and obtained utility. Our theory of real price depends on the asymptotic analysis of a Markov chain transition probability matrix related to the network of producers and consumers. We show that the real price of a product or service can be determined from the involved Markov chain matrix, and can be dramatically different from the product's label price. In particular, we show that the label prices of products and services are often not ``real'' or directly ``useful'': given two products offering the same myopic utility, the one with lower label price may not necessarily offer better asymptotic utility. This theory shows that the globality or locality of the  products and services does have different impacts on the spending-utility tradeoff of a customer. The established mathematical theory of real price can be used to determine whether to adopt or not to adopt certain artificial intelligence (AI) technologies from an economic perspective.

\end{abstract}


\section{Introduction}


 The world's economy has become increasingly interconnected, due to the evolution of global investments, trades and e-commerce activities \cite{stiglitz2017overselling,hummels2007transportation, freund2004effect}. These global economic activities have become an integrated part of modern society.  On the other hand, there are also ideas advocating the practice of ``buy local'', by which people buy locally produced goods and services rather than those produced farther away. ``Buy local'' sometimes refers to buying locally grown agricultural produces, but it can also includes buying other locally produced products and services.

Why do people buy locally? Recent studies have focused on moral arguments or ethical concerns and yet led to different conclusions. For example, on the one hand, \cite{ferguson2021buy} argues that buying local food is morally superior to the alternatives because it helps the poorest food workers economically while also mitigating the environmental impact of food transportation. On the other hand, \cite{young2022should} analyzes the main justifications of locavorism in the existing literature, namely protecting the environment, promoting community, promoting small business, and contributing to the prosperity of one’s local economy, and show that none of them provides a strong positive ethical reason for consumers in general to adopt the practice of buying local.

However, only a few papers have formally explored the economic implications of ``buy local" decisions in a generalized theoretical model. For example, \cite{winfree2017welfare} provides a model to show that market frictions such as economic externalities and market power considerations can explain the decision of buying local even when the region has no competitive advantage. But in their model, buying local may still lead to deadweight losses in the absence of market power or externalities.  In addition, \cite{mailath2016buying} provides a formal model with a monopolistically competitive economy that is mainly built on reciprocity and trust to explain ``buy local". However, like other existing literature, they focus on the equilibrium outcomes in a short-term transactions.

In this paper, we argue that one needs not appeal to social factors or environmental considerations to motivate ``buy local" decisions and that buying locally can be beneficial even when agents are motivated strictly by selfish concerns.  Our studies are based on long-term analysis of the interconnected producer-consumer network from a theoretical perspective. 

In particular, we establish a mathematical theory of real price that determines the optimal global versus local spending of an agent which achieves the agent's optimal tradeoff between spending and utility. Essentially, the real price is the equivalent hypothetical price one customer has to pay for one unit of product such that the customer obtains the same myopic utility as the true asymptotic average utility. Our theory of real price depends on the asymptotic analysis of a Markov chain transition probability matrix related to the network of producers and consumers. We show that the real price of a product or service can be determined from the involved Markov chain matrix via solving an optimization problem involving the stationary distribution of the underlying Markov chain. The real price can be dramatically different from the product's label price. In particular, we show that the label prices of products and services are often not ``real'': given two products offering the same myopic utility, the one with lower label price may not necessarily offer better asymptotic utility. This theory shows that the globality or locality of the  products and services of the same label prices may have different impacts on the spending and utility tradeoff.

The established mathematical theory of real price can also be used to determine whether to adopt or not to adopt certain artificial intelligence (AI) technologies from an economic perspective. AI is very promising in further automating many tasks that used to be performed fully or partially by humans \cite{rosenblatt1961principles}. In general, automation via AI technologies can greatly increase the overall production efficiency of the society. For example, performing a certain task using AI can come at a much lower label price than other means of performing the same task. However, our theory shows that, even though the AI technologies may come at a lower label price, these AI technologies can carry a higher ``real '' price than their non-AI alternatives, for communities adopting them. From an economic perspective, the benefits of adopting AI technologies depend on the structure of the network of technology providers and technology consumers, and on the ``locality'' of these AI technologies to the communities adopting them.

This paper is organized as follows. In Section \ref{Sec:mathmodel}, we establish a Markov chain based mathematical model to investigate the problem of optimizing local and global spending. In Section \ref{optimization}, we formulate a polynomial optimization problem to optimize the spending patterns of agents to achieve their optimal spending and utility tradeoff. In Section \ref{sec:realprice}, we give the concept of real prices, and derive the formulas for real prices in the context of a spending network.  In Section \ref{numerical}, we present numerical results demonstrating that our optimized spending strategy can significantly outperform direct label-price based spending strategy. In Section \ref{conclusions}, we conclude our paper and lay out future investigation directions.


\section{Mathematical Model}\label{Sec:mathmodel}
We consider $n$ agents, where $n$ is a positive integer. Each agent can function as a producer and also as a consumer. We consider a spending matrix $P\in \mathbb{R}^{n\times n}$, where $P$ is a Markov chain transition probability matrix. We denote the element in the $i$-th row and $j$-th column of $P$ as $P_{i,j}$. For matrix $P$, we have
$\sum_{i=1}^{n}P_{i,j}=1$ for every $1 \leq j \leq n$. The number $P_{i,j}$ denotes the fraction of the $j$-th agent's currency the $j$-th agent spends on the services and products provided by the $i$-th agent. 

We use $x^t \in \mathbb{R}^{n\times 1}$ to denote the currency vector at time index $t$, where $t\geq 0$ is an integer.  We assume that each element of $x^t$ is non-negative, representing the possessed currency by each agent at time index $t$. We refer to each time index as one episode.  We also have a utility matrix  $U\in \mathbb{R}^{n\times n}$ whose elements are non-negative. We denote the element in the $i$-th row and $j$-th column of $U$ as $U_{i,j}$, which represents the utility the $i$-th agent provides to the $j$-th agent if the $j$-th agent purchases one unit of product or service from Agent $i$. We further have a price matrix $C \in \mathbb{R}^{n\times n}$ such that $C_{i,j}$, the element in the $i$-th row and $j$-th column of $C$, represents the price that the $i$-th agent charges for providing a unit of service or product to the $j$-th agent.  We assume that no agent changes their spending pattern as time evolves, namely an agent maintains the relative ratios of spending on each agent.

 On the surface, to maximize the utility of the $j$-th agent, the $j$-th agent should spend all its currency on the agent who provides the lowest price and largest utility per unit product, namely Agent $i$ with the largest $\frac{U_{i,j}}{C_{i,j}}$. However, counter-intuitively we argue that this is not true, and the optimal spending practice should depend on the structure of the network of producers and consumers. In fact, we will give an optimization formulation in the next section which provides the optimal spending pattern for each agent.

We notice that at time index $(t+1)$,  the currency vector is updated to 
$$ x^{t+1}=Px^{t}.$$

\section{Optimization of Spending Patterns}
\label{optimization}

Without loss of generality, we focus on the $1$-st agent and try to determine its optimal spending strategy, namely determining the first column of matrix $P$. Our goal is to maximize the utility obtained by Agent $1$ over time.

We assume that the matrix $P$ is an irreducible Markov chain transition matrix. A sufficient condition, denoted by $``CD''$,  for $P$ to be irreducible is that both of the following two conditions hold: (1) for each $j$, $1\leq j \leq n$,  we have the submatrix $P_{/j,/j}$ is irreducible, where $P_{/j,/j}$ is the submatrix of $P$ excluding the $j$-th row and $j$-th column; (2) for each $j$, $1\leq j \leq n$, there exists an agent $k\neq j$ such that $P_{j,k}>0$ and there exists an agent $l\neq k$ such that $P_{l,j}>0$.  

Suppose that the matrix $P$ satisfies the condition $``CD''$. From $x^{t+1}=Px^{t}$, the sum of the elements of the currency vector remains constant as time index $t$ increases. By scaling $x^t$ to $(x')^t$ such that the sum of the elements of $(x')^t$ is equal to 1, we can view the vector $(x')^t$ as a probability distribution vector. We can view $P$ as the transition probability matrix of a Markov chain.  Then because of irreducibility of this Markov chain, the Markov chain will have a unique stationary distribution. If this Markov chain is further aperiodic, $(x')^t$ will converge to this unique stationary probability distribution vector $x'$ satisfying $Px'=x'$. Then the currency vector $x$ will converge to a unique stationary currency vector $x$ satisfying $Px=x$. Even if the Markov chain is not aperiodic, by the ergodic theorem for irreducible Markov chains, we know the proportion of the time the Markov chain spends on each state converges to that unique stationary distribution as time index increases, with high probability, no matter what the starting distribution $(x')^0$ is.  So when the Markov chain is irreducible, we can always use its unique stationary distribution to calculate the asymptotic utility per episode.

In order to maximize the utility of the $j$-th agent under the stationary currency vector, we formulate the following optimization problem: 


\begin{align}\label{Defn:utilitymaximization}
&\underset{x,~P_{:,j}}{\rm minimize\ }  x_j\sum_{i=1}^{n} \frac{P_{i,j} U_{i,j}}{C_{i,j}}\nonumber\\
&{\rm subject\ to\ }~~~~~x\geq 0, \nonumber \\
&~~~~~~~~~~~~~~~~~~~Px=x, \nonumber \\
&~~~~~~~~~~~~~~~~~~~\|x\|_1=\|x^{0}\|_1,\nonumber\\
&~~~~~~~~~~~~~~~~~~~P{:,j}\geq 0.
\end{align}
 where $P_{:,j}$ is the $j$-th column of $P$, and $\|x\|_1$ represents the $\ell_1$ norm of $x$, namely the sum of the absolute values of elements in $x$. For simplicity, we enforce that the elements of $x$ be non-negative.   In this case, the vector $x$ represents the scaled stationary distribution of the probability transition matrix $P$. Because the system reaches the scaled stationary probability distribution, we must have the second constraint $Px=x$. Because the sum of the elements of $x$ does not change over time, we have $\|x\|_1=\|x^{0}\|_1$. Since $x_j$ is the stationary amount of currency agent $j$ will possess in the stationary distribution, we have the objective function in (\ref{Defn:utilitymaximization}).

 Without loss of generality, we can assume that $\|x^{0}\|_1=1$.  Under this assumption, we can change (\ref{Defn:utilitymaximization}) to : 
\begin{align}\label{unit_utilitymaximization}
&\underset{x,~P_{:,j}}{\rm minimize\ }  x_j\sum_{i=1}^{n} \frac{P_{i,j} U_{i,j}}{C_{i,j}}\nonumber\\
&{\rm subject\ to\ }~~~~~x\geq 0, \nonumber \\
&~~~~~~~~~~~~~~~~~~~Px=x, \nonumber \\
&~~~~~~~~~~~~~~~~~~~\|x\|_1=1,\nonumber\\
&~~~~~~~~~~~~~~~~~~~P{:,j}\geq 0.
\end{align}

We observe that (\ref{unit_utilitymaximization}) is a non-linear optimization problem since the objective function involves products of elements of $x$ and elements of $P$.
To numerically solve (\ref{unit_utilitymaximization}),  we can consider  (\ref{unit_utilitymaximization}) as a polynomial optimization problem, transform it into a sum of squares optimization formulation \cite{Parrilo2004SumOS} and related semidefinite programming, and use algorithms for semidefinite programming to find the optimal $x$ and spending strategy (allocation) $P_{:,j}$.

 We further observe that (\ref{unit_utilitymaximization}) becomes a linear programming problem in the other elements of $x$ and the elements of $P$, if we fix a certain value for $x_1$. This observation leads to the following numerical algorithm which avoids using semidefinite programming for polynomial optimization: the semidefinite programming can be time-consuming for large $n$.  In our algorithm, we solve (\ref{unit_utilitymaximization}) by searching over possible values of $x_1$ (for example, grid search). For each examined value of $x_1$, we solve the corresponding linear programming problem. Then among all the values for $x_1$, we select the one which gives the highest objective function value.  Because we only do grid search (or other types of search) over one-dimensional variable $x_1$, and $x_1$ is bounded between $0$ and $1$, this algorithm can work with relatively low complexity. 

\section{Real Price Versus Label Price}
\label{sec:realprice}
In this section, we give the definition of real price of Agent $k$'s products and services seen from Agent $j$ under a given spending pattern. Essentially Agent $i$ spends a certain amount of currency purchasing products and services from Agent $k$ at this real price will obtain an instant utility (within one single time episode) that is equal to ``the asymptotic utility per episode if Agent $j$ spends the same amount of currency on Agent $k$'s products and services at the label price.''


\subsection{Real Price under Fixed Spending Pattern}
Let us define the fixed spending pattern of agent $j$ as a spending pattern where agent $j$ allocates a fixed percentage of spending on the products and services provided by each agent. From the formulation above, we can see that the system reaches a stationary distribution if the involved Markov chain is irreducible.  

Suppose that agent $j$ spends an additional $\Delta a$ on purchasing the products and services provided by agent $k$ at the initial time index ($t=0$), while maintaining a fixed spending pattern (meaning that agent $j$'s spending on products and services provided by other agents is also increased proportionally). 

We know the asymptotic utility per episode under a fixed spending pattern is given by 
$$W=\sum_{i=1}^{n} \|x^{0}\|_1 \frac{x_j U_{i,j}}{C_{i,j}},$$
where $x$ is the stationary distribution vector. 

Thus the increase in the the asymptotic utility per episode due to the additional spending of agent $j$ on agent $k$ is given by 

$$\Delta W=\sum_{i=1}^{n} (x_j \Delta a)  \frac{ U_{i,j}}{C_{i,j}}.$$

This additional asymptotic utility per episode is equivalent to the utility agent $j$ will obtain in the $0$-th episode by spending $\Delta a$ directly on agent $k$'s products and services with a new price $X(\Delta a)$ (as a function of $\Delta a$), where 
$$\sum_{i=1}^{n} (x_j \Delta a)  \frac{ U_{i,j}}{C_{i,j}}= \frac{\Delta a U_{k,j}}{X(\Delta a)}.$$
Solving for $X(\Delta a)$, we obtain 
$$X(\Delta a)=  \frac{U_{k,j}}{\sum_{i=1}^{n} x_j \frac{ U_{i,j}}{C_{i,j}} }. $$

We call $X(\Delta a)$ the real price.  It is equivalent to the marginal price often see in economics, but here it is obtained from an asymptotic analysis of the Markov chain. 

\begin{defn} (Real Price of agent $k$'s products and services when agent $j$ makes purchases from agent $k$)
Suppose that agent $j$ spends an additional $\Delta a$ currency on purchasing the products and services provided by agent $k$ at the initial time index ($t=0$), and suppose that the involved matrix chain is an irreducible Markov chain. Suppose that the asymptotic utility per episode as $t\rightarrow \infty$ increases by $\Delta W$. We define the real price of agent $k$'s products and services seen by agent $j$ as 
$$RP(\Delta a)= \frac{U_{k,j}\Delta a} {\Delta W}. $$
We further define the instant real price as 
$$RP= \lim_{\Delta a \rightarrow 0}\frac{U_{k,j}\Delta a} {\Delta W}. $$

\end{defn}

For example, we can see that, when we maintain agent $j$'s fixed spending ratio, the real price of agent $k$'s products and services seen from agent $j$ is given by  
$$RP=\lim_{\Delta a \rightarrow 0} X(\Delta a) = \frac{1}{x_j} \frac{U_{k,j}}{\sum_{i=1}^{n} \frac{ U_{i,j}}{C_{i,j}} }.$$
This is in general different from the label price of each agent's products and services. In fact, the agents offering lower label prices can correspond to higher real prices, meaning it is in fact not desirable to purchase from the agents offering lower label price.

 \subsection{Real price under dynamic spending pattern}
 \label{realprice_dynmaic}
In this subsection, we consider the case where agent $j$ singularly spends an additional $\Delta a$ currency on purchasing agent $k$'s products and services at $t=0$. Here, we do not increase the amount of currency spent by agent $j$ on other agents' products and services. Without loss of generality, we assume that $j=1$. We denote the total currency as $T=\|x^0\|_1$ (before the spending increases), and the new total amount of currency (after the spending increases) as $T'=\|x^0\|_1+\Delta a$.  This new addition will results in a new spending pattern for Agent $j$ at $t=0$. We assume that as time index $t$ increases, Agent $j$ keeps their new initial spending pattern at $t=0$, even as time index $t$ increases.

\subsubsection{Number of agents $n=3$}
  We consider the case $n=3$, and will later extend it to general $n$. As $t \rightarrow \infty$, suppose that we have arrived at the stationary currency vector $x$. Namely, we have

$$\begin{pmatrix}
    \frac{a_1}{a_1+a_2+a_3} & P_{1,2} & P_{1,3}\\
    \frac{a_2}{a_1+a_2+a_3} & P_{2,2} & P_{2,3}\\
    \frac{a_3}{a_1+a_2+a_3} & P_{3,2} & P_{3,3}
\end{pmatrix}  
\begin{pmatrix}
    x_1 \\
    x_2 \\
    x_3
\end{pmatrix}
=
 \begin{pmatrix}
    x_1 \\
    x_2 \\
    x_3
\end{pmatrix},
\label{eq:stationaryRPD}
$$
and 
$$x_1+x_2+x_3= \|x^0\|_1,$$
where $a_1$, $a_2$, and $a_3$ are the amounts of currency agent $j$ spends on each agent's products and services at $t=0$, and $a_1+a_2+a_3=x^0_1$.

Now we change $a_2$ to $a_2+\Delta a_2$. Thus we have $$x_1+x_2+x_3=\|x^0\|_1+ \Delta a_2,$$
and 
$$a_1+a_2+a_3=x^0_1+ \Delta a_2.$$

Suppose that the change in the total utility per episode for agent $j=1$ is $\Delta W$. We aim to find the instant real price 
$$RP= \lim_{\Delta a_2 \rightarrow 0}\frac{U_{k,j}\Delta a_2} {\Delta W},$$
where $k=2$ and $j=1$.

First of all, we have the following relations as $\Delta a_2 \rightarrow 0$ by taking the total derivative of euqations for the stationary distribution:
\[\Delta{}x_1+\Delta{}x_2+\Delta{}x_3=\Delta{}a_2\]

\[-\frac{a_1}{(a_1+a_2+a_3)^2}\Delta{}a_2x_1+\frac{a_1}{(a_1+a_2+a_3)}\Delta{}x_1+P_{1,2}\Delta{}x_2+P_{1,3}\Delta{}x_3=\Delta{}x_1\]

\[[\frac{1}{a_1+a_2+a_3}-\frac{a_2}{(a_1+a_2+a_3)^2}]\Delta{}a_2x_1+\frac{a_2}{a_1+a_2+a_3}\Delta{}x_1+P_{2,2}\Delta{}x_2+P_{2,3}\Delta{}x_3=\Delta{}x_2\]

\[-\frac{a_3}{(a_1+a_2+a_3)^2}\Delta{}a_2x_1+\frac{a_3}{a_1+a_2+a_3}\Delta{}x_1+P_{3,2}\Delta{}x_2+P_{3,3}\Delta{}x_3=\Delta{}x_3\]

Reorganizing, we have the following system of equations:  

\[\Delta{}x_1+\Delta{}x_2+\Delta{}x_3=\Delta{}a_2\]

\[\left(\frac{a_1}{a_1+a_2+a_3}-1\right)\Delta x_1+P_{1,2}\Delta{}x_2+P_{1,3}\Delta{}x_3= \frac{a_1}{(a_1+a_2+a_3)^2}x_1 \Delta a_2\]

\[\frac{a_2}{a_1+a_2+a_3}\Delta{}x_1+(P_{2,2}-1)\Delta{}x_2+P_{2,3}\Delta{}x_3= -[\frac{1}{a_1+a_2+a_3}-\frac{a_2}{(a_1+a_2+a_3)^2}] x_1 \Delta a_2 \]

\[\frac{a_3}{a_1+a_2+a_3}\Delta{}x_1+P_{3,2}\Delta{}x_2+(P_{3,3}-1)\Delta{}x_3=\frac{a_3}{(a_1+a_2+a_3)^2} x_1 \Delta a_2.\]
One of the four equations is redundant because in the Markov chain transition probability matrix $P$, the elements in each column sum up to $1$. From the four equations above, we can determine the the ratios 
$\frac{\Delta x_1}{\Delta a_2}$, $\frac{\Delta x_2}{\Delta a_2}$, and $\frac{\Delta x_3}{\Delta a_2}$. We further have


\begin{align}
\Delta W&={\sum_{i=1}^{3}\Delta{}\left(x_1 \frac{a_i}{a_1+a_2+a_3}\frac{U_{i,1}}{C_{i,1}}\right)}\\
&=\sum_{i=1}^{3}\left[\frac{U_{i,1}}{C_{i,1}}\Delta{}\left(x_1\frac{a_i}{a_1+a_2+a_3}\right)\right]\\
&=\sum_{i=1}^{3}\frac{U_{i,1}}{C_{i,1}}\left[\Delta{}x_1\frac{a_i}{a_1+a_2+a_3}+x_1\Delta{}\left(\frac{a_i}{a_1+a_2+a_3}\right)\right] \\
&=\sum_{i=1, i\neq 2}^{3}\frac{U_{i,1}}{C_{i,1}}\left[\Delta{}x_1\frac{a_i}{a_1+a_2+a_3}-x_1\frac{a_i}{(a_1+a_2+a_3)^2}\Delta{}a_2\right]\\
&~~+ \frac{U_{i=2,1}}{C_{i=2,1}}\left[\Delta{}x_1\frac{a_i}{a_1+a_2+a_3}+x_1\left[\frac{1}{a_1+a_2+a_3}\Delta{}a_2-\frac{a_2}{(a_1+a_2+a_3)^2}\Delta{}a_2\right]\right]
\end{align}

Then the instant real price is given by: 
$$RP= \lim_{\Delta a_2 \rightarrow 0}\frac{U_{k,j}\Delta a_2} {\Delta W},$$
which can be calculated since we know $\frac{\Delta x_1}{\Delta a_2}$, $\frac{\Delta x_2}{\Delta a_2}$, and $\frac{\Delta x_3}{\Delta a_2}$.


\subsubsection{General number of agents $n$}
\label{subsection_generaln}
We consider a general $n$. As $t \rightarrow \infty$, we have arrived at the stationary currency vector $x$. Namely, we have

$$\begin{pmatrix}
    \frac{a_1}{\sum_{i=1}^{n}a_i} & P_{1,2} & P_{1,3} & \dots{} & P_{1,n}\\
    \frac{a_2}{\sum_{i=1}^{n} a_i} & P_{2,2} & P_{2,3} & \dots{} & P_{2,n}\\
    \vdots{} & \vdots{} & \vdots{} & \dots{} & \vdots{}\\
    \frac{a_n}{\sum_{i=1}^{n} a_i} & P_{n,2} & P_{n,3} & \dots{} & P_{n,n}
\end{pmatrix}  
\begin{pmatrix}
    x_1 \\
    x_2 \\
    \vdots{} \\ 
    x_n
\end{pmatrix}
=
 \begin{pmatrix}
    x_1 \\
    x_2 \\
    \vdots{} \\ 
    x_n
\end{pmatrix},
\label{eq:stationaryRPD}
$$
and 
$$x_1+x_2+\dots{}x_n= \|x^0\|_1,$$
where $a_1$, $a_2$, $\dots$, $a_n$ are the amount of currency agent $j$ spends on each agent's products and services at $t=0$, and $a_1+a_2+\dots+a_n=x^0_1$.

Now we change $a_2$ to $a_2+\Delta a_2$,  and after the change, we thus have $$x_1+x_2+\dots{}+x_n=\|x^0\|_1+ \Delta a_2,$$
and 
$$a_1+a_2+\dots{}+a_n=x^0_1+ \Delta a_2.$$

Suppose that the change in the total utility per episode for agent $j=1$ is $\Delta W$, we aim to find the instant real price 
$$RP= \lim_{\Delta a_2 \rightarrow 0}\frac{U_{k,j}\Delta a_2} {\Delta W},$$
where $k=2$ and $j=1$. Note that if $j\neq 1$ or $k\neq 2$, one can always do column or row exchanges of matrix $P$ such that the agents of interest take indices $k=2$ and $j=1$.

First of all, we have the following relations as $\Delta a_2 \rightarrow 0$ by taking the total derivative:
\[\Delta{}x_1+\Delta{}x_2+\dots{}+\Delta{}x_n=\Delta{}a_2\]

\[-\frac{a_1}{(\sum_{i=1}^{n} a_i)^2}\Delta{}a_2x_1+\frac{a_1}{\sum_{i=1}^{n} a_i}\Delta{}x_1+\sum_{i=2}^{n}P_{1,i}\Delta{}x_i=\Delta{}x_1\]

\[[\frac{1}{\sum_{i=1}^{n}a_i}-\frac{a_2}{(\sum_{i=1}^{n}a_i)^2}]\Delta{}a_2x_1+\frac{a_2}{\sum_{i=1}^{n} a_i}\Delta{}x_1+\sum_{i=2}^{n}P_{2,i}\Delta{}x_i=\Delta{}x_2\]

\[-\frac{a_3}{(\sum_{i=1}^{n} a_i)^2}\Delta{}a_2x_1+\frac{a_3}{\sum_{i=1}^{n} a_i}\Delta{}x_1+\sum_{i=2}^{n}P_{3,i}\Delta{}x_i=\Delta{}x_3\]

\[\vdots{}\]

\[-\frac{a_n}{(\sum_{i=1}^{n} a_i)^2}\Delta{}a_2x_1+\frac{a_n}{\sum_{i=1}^{n} a_i}\Delta{}x_1+\sum_{i=2}^{n}P_{n,i}\Delta{}x_i=\Delta{}x_n\]

Reorganizing, we have the following system of equations:  

\[\Delta{}x_1+\Delta{}x_2+\dots{}+\Delta{}x_n=\Delta{}a_2\]

\[\left(\frac{a_1}{\sum_{i=1}^{n} a_i}-1\right)\Delta x_1+\sum_{i=2}^{n}P_{1,i}\Delta{}x_i= \frac{a_1}{(\sum_{i=1}^{n} a_i)^2}x_1 \Delta a_2\]

\[\frac{a_2}{\sum_{i=1}^{n} a_i}\Delta{}x_1-P_{2,2}\Delta{}x_2+\sum_{i=2}^{n}P_{2,i}\Delta{}x_i = -[\frac{1}{\sum_{i=1}^{n} a_i}-\frac{a_2}{(\sum_{i=1}^{n} a_i)^2}] x_1 \Delta a_2 \]

\[\frac{a_3}{\sum_{i=1}^{n} a_i}\Delta{}x_1-P_{3,3}\Delta{}x_3+\sum_{i=2}^{n}P_{3,i}\Delta{}x_i=\frac{a_3}{(\sum_{i=1}^{n} a_i)^2} x_1 \Delta a_2.\]

\[\vdots{}\]

\[\frac{a_n}{\sum_{i=1}^{n} a_i}\Delta{}x_1-P_{n,n}\Delta{}x_n+\sum_{i=2}^{n}P_{n,i}\Delta{}x_i=\frac{a_n}{(\sum_{i=1}^{n}a_i)^2} x_1 \Delta a_2.\]
One of the $n+1$ equations is redundant because the in the Markov chain transition probability matrix $P$, the elements of each column sum up to $1$. From the $n+1$ equations above, we can determine the the ratios 
$\frac{\Delta x_1}{\Delta a_2}$, $\frac{\Delta x_2}{\Delta a_2}$, $\dots$, $\frac{\Delta x_n}{\Delta a_2}$. 



Thus the change in the utility in each episode for agent $j=2$ is equal to:
\begin{align}
\Delta W&={\sum_{i=1}^{n}\Delta{}\left(x_1 \frac{a_i}{\sum_{i=1}^{n} a_i}\frac{U_{i,1}}{C_{i,1}}\right)}\\
&=\sum_{i=1}^{n}\left[\frac{U_{i,1}}{C_{i,1}}\Delta{}\left(x_1\frac{a_i}{\sum_{i=1}^{n} a_i}\right)\right]\\
&=\sum_{i=1}^{n}\frac{U_{i,1}}{C_{i,1}}\left[\Delta{}x_1\frac{a_i}{\sum_{i=1}^{n} a_i}+x_1\Delta{}\left(\frac{a_i}{\sum_{i=1}^{n} a_i}\right)\right] \\
&=\sum_{i=1, i\neq 2}^{n}\frac{U_{i,1}}{C_{i,1}}\left[\Delta{}x_1\frac{a_i}{\sum_{i=1}^{n} a_i}-x_1\frac{a_i}{(\sum_{i=1}^{n}a_i)^2}\Delta{}a_2\right]\\
&~~+ \frac{U_{i=2,1}}{C_{i=2,1}}\left[\Delta{}x_1\frac{a_i}{\sum_{i=1}^{n} a_i}+x_1\left[\frac{1}{\sum_{i=1}^{n} a_i}\Delta{}a_2-\frac{a_2}{(\sum_{i=1}^{n} a_i)^2}\Delta{}a_2\right]\right]
\end{align}

Then the instant real price is given by: 
$$RP= \lim_{\Delta a_2 \rightarrow 0}\frac{U_{k,j}\Delta a_2} {\Delta W},$$
which can be calculated since we know $\frac{\Delta x_1}{\Delta a_2}$, $\frac{\Delta x_2}{\Delta a_2}$, $\dots$,  $\frac{\Delta x_n}{\Delta a_2}$.

\section{Numerical Results}
\label{numerical}

\begin{figure}
\begin{center}
\includegraphics[width=\textwidth]{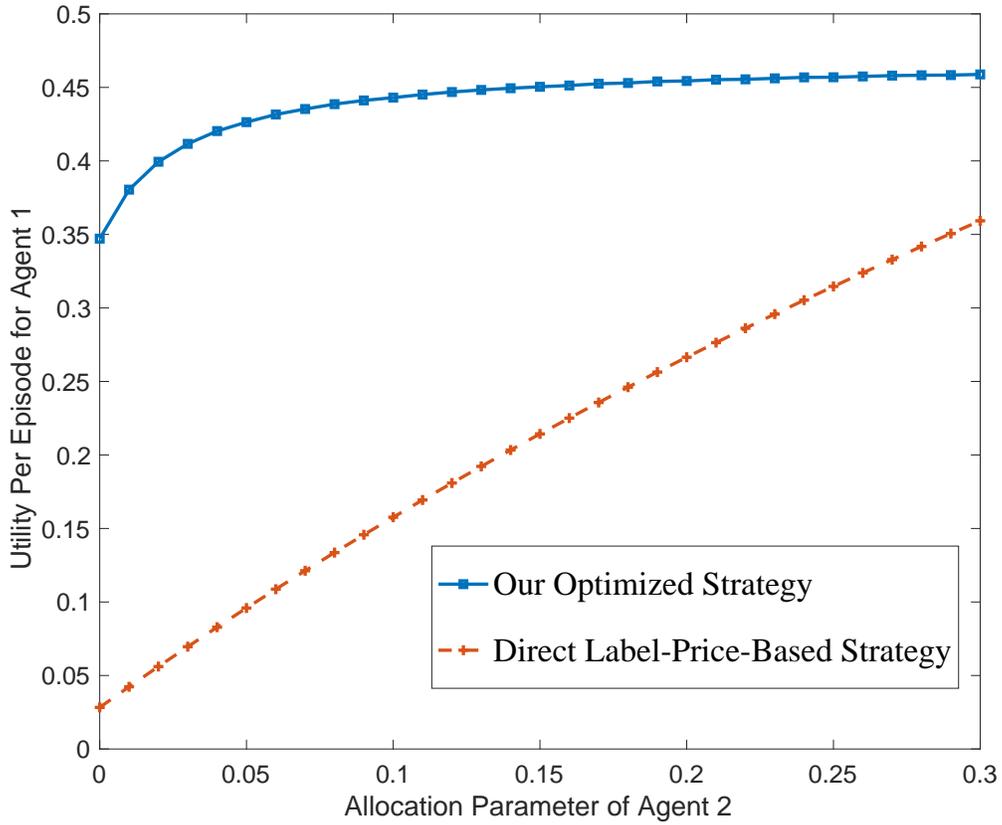}
\caption{Comparisons in terms of utility per episode obtained by Agent 1, between through our optimal spending strategy for Agent 1, and through direct label-price-based strategy for Agent 1. This is done in a 3-agent network of producers and consumers. The x-axis is $\alpha$. }
\label{fig:utilitycomparison}
\end{center}
\end{figure}

\begin{figure}
\begin{center}
\includegraphics[width=\textwidth]{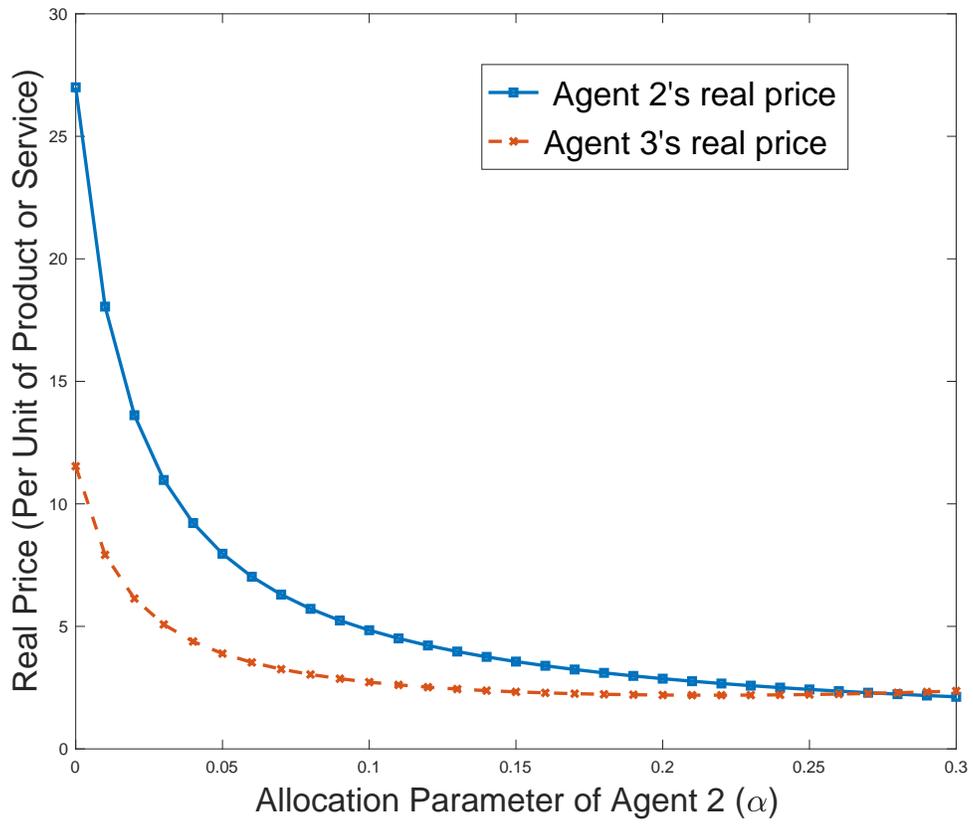}
\caption{Comparisons of real prices from Agent 2 and 3, seen from the perspective of Agent 1. This is done in a 3-agent network of producers and consumers. The x-axis is $\alpha$. }
\label{fig:realprice}
\end{center}
\end{figure}

\begin{figure}
\begin{center}
\includegraphics[width=\textwidth]{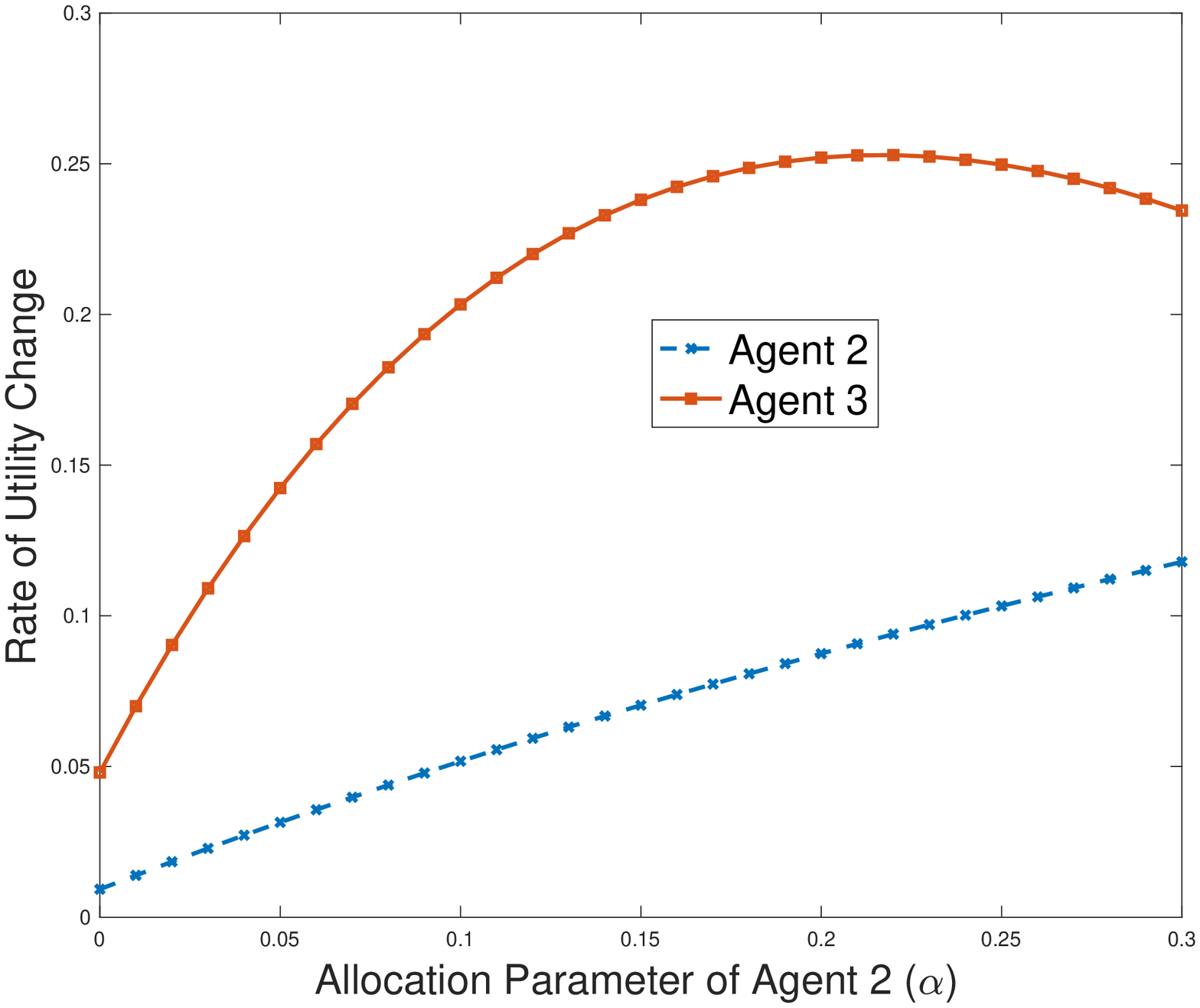}
\caption{Comparisons of rates of change for the asymptotic utility (namely marginal asymptotic utility) obtained by Agent 1, with respect to the change in $a_2$ (Agent 2) and $a_3$ (Agent 3). This is done in a 3-agent network of producers and consumers. The x-axis is $\alpha$. }
\label{fig:rateofchange}
\end{center}
\end{figure}

We consider a 3-agent network for simplicity (even though the codes work for general $n$). In this network, we focus on optimizing the spending strategy for Agent 1, give that Agents 2's and 3's spending strategies are fixed.  We consider the following Markov chain matrix:
$$
P=\begin{pmatrix}
    P_{1,1} & P_{1,2} & P_{1,3}\\
    P_{2,1} & P_{2,2} & P_{2,3}\\
    P_{3,1} & P_{3,2} & P_{3,3}
\end{pmatrix},  
$$
where we need to optimize $P_{1,1}$, $P_{2,1}$ and $P_{3,1}$.  In our experiment, we adopt the following numerical values for the last two columns of $P$:
$$
P=\begin{pmatrix}
    P_{1,1} & \alpha & 0.5\\
    P_{2,1} & 1-\alpha-0.02 & 0.01\\
    P_{3,1} & 0.02 & 0.49
\end{pmatrix},  
$$
where $\alpha$ is a tunable parameter for Agent 2. The larger $\alpha$ is, more willing is Agent 2 to spend their currency on purchasing Agent 1's products and services. 

\subsection{Comparisons between optimized spending strategy against direct label price based strategy}
 We (randomly) generated the following utility matrix $U$ and price matrix $C$:
$$
U=\begin{pmatrix}
    0 & 0.3746 & 0.6199\\
    0.7637 & 0 & 1.0246\\
    0.5495 & 0.3102 & 0
\end{pmatrix},  
$$
and
$$
C=\begin{pmatrix}
0.1619  &  6.7064  &  0.3463\\
    0.5000 &   4.6920  &  6.0502\\
    0.3958 &   0.7562  &  7.1731
\end{pmatrix}.
$$

Under such $U$ and $C$, Agent 2 will provide the highest utility-over-price ratio for Agent 1. So under the myopic approach, Agent 1 would spend all its currency on purchasing products and services from Agent 2.  Contrary to the myopic approach, we will find the optimized spending strategy for Agent 1, via solving the optimization problem (\ref{unit_utilitymaximization}). Figure \ref{fig:utilitycomparison} shows the utility per episode obtained by Agent 1 as a function of $\alpha$, comparing our optimal spending strategy for Agent 1, against the direct label-price-based strategy (picking the producer offering best utility-over-price ratio) for Agent 1. 

As shown by Figure \ref{fig:utilitycomparison}, our optimized spending strategy can provide more than 10 fold improvement in asymptotic utility per episode for Agent 1 under small $\alpha$ values.  When $\alpha$ is small, Agent 1 can obtain higher asymptotic utility per episode by spending majority of its currency on Agent 3's products and services, even though Agent 3 provides a lower direct utility-over-price ratio. The intuition for this phenomenon is that the currency (of Agent 1) spent on Agent 2's products and services has a lower proportion coming back to purchasing Agent 1's products and services so in the stationary distribution of currency, Agent 1 will have less money that can be used for purchasing,  namely a smaller $x_2$. An acute reader might find Agent 2 applies a portion of its currency on itself, but this is more about the modeling technique. In reality, Agent 2 can be spending that portion on the products and services of another outside agent (say, Agent 4), and that outside agent spends all its currency on Agent 2's products and services. In such a case, the modeling is equivalent to the 3-agent network modeling we described, as if Agent 2 is spending that portion of currency on itself. 

As $\alpha$ grows, Agent 2 is more willing to spend currency purchasing products and services from Agent 1, and the gap between the myopic approach and our optimized approach becomes smaller.

\subsection{Computations of Real Prices}

In this subsection, we consider the same problem setup described at the beginning of the numerical section.  However, for this subsection, we are using a new utility matrix $U$ and a price matrix $C$:

$$
U=\begin{pmatrix}
         0  &  0.7589  &  0.5426\\
    0.2507  &       0  &  1.1631\\
    0.5542  &  0.2726   &      0
\end{pmatrix},  
$$
and
$$
C=\begin{pmatrix}
    0.6380  &  0.1850  &  0.4736\\
    0.5000  &  2.6813  &  0.9683\\
    1.2161  &  2.5220  &  1.5043
\end{pmatrix}.  
$$

We focus on finding the real prices under dynamic spending pattern, namely Agent 1 does not proportionally increase spending on the other agents.  
For the computations, we assume that $a_1=0$, $a_2=2$, $a_3=0$, and  $\|x^0\|_1=10$. We further assume that the  matrix $P$ is given as follows:
$$
P=\begin{pmatrix}
    \frac{a_1}{a_1+a2+a_3} & \alpha & 0.5\\
    \frac{a_2}{a_1+a2+a_3} & 1-\alpha-0.02 & 0.01\\
    \frac{a_3}{a_1+a2+a_3} & 0.02 & 0.49
\end{pmatrix},  
$$
where $\alpha$ is a tunable parameter for Agent 2. 

Figure \ref{fig:realprice} plots the real prices of Agents 2 and 3's products and services, as seen from the perspective of Agent 1. As we can see, even though the label price of Agent 2's products and services is lower than that of Agent 3, Agent 2's real price is much higher.


In Figure \ref{fig:rateofchange}, We also plot the rate of change for the asymptotic utility per episode obtained by Agent 1, with respect to the change in $a_2$ and $a_3$. The x-axis represents the parameter $\alpha$ in the model. As we can see,  changing $a_3$ (increase spending on Agent 3) has a more significant positive effect on the utility obtained by Agent 1, even though Agent 3's products and services are of higher prices and have a lower utility-over-price ratio.

From these numerical results, we can see that sometimes buying from local providers (those agents who may have higher interest in purchasing from Agent 1) can potentially boost the asymptotic utility for Agent  1, even though the myopic prices and utilties form local providers look inferior ostensibly.

\section{Conclusions and Future Directions}
\label{conclusions}
In this paper, we establish a theoretical model to determine whether buying locally or adopting certain AI technologies is worthwhile. Our model is based on asymptotic analysis of associated Markov chains of agents' spending patterns, and our optimized spending strategy is obtained by optimization of the Markov chain's transition probabilities. Surprisingly, we show that the label prices of products and services are often not indicative of whether these products and services are worthwhile for purchase. Our model presents an interesting mathematical optimization problem involving a Markov chain. This formulation can potentially be applied to many areas, and can be extended in many possible ways. For example, one can consider general utilities as non-linear functions of spending, consider more finely defined products and services, dynamic total currency amount, competitions and collaborations among agents, competitive pricing for products and services, profits maximization for agents, and time value of money.    

\bibliography{Reference_NullSpaceCondition.bib}

\begin{thebibliography}{1}

\bibitem{ferguson2021buy}
Benjamin Ferguson and Christopher Thompson.
\newblock Why buy local?
\newblock {\em Journal of Applied Philosophy}, 38(1):104--120, 2021.

\bibitem{freund2004effect}
Caroline~L Freund and Diana Weinhold.
\newblock The effect of the internet on international trade.
\newblock {\em Journal of international economics}, 62(1):171--189, 2004.

\bibitem{hummels2007transportation}
David Hummels.
\newblock Transportation costs and international trade in the second era of
  globalization.
\newblock {\em Journal of Economic perspectives}, 21(3):131--154, 2007.

\bibitem{mailath2016buying}
George~J Mailath, Andrew Postlewaite, and Larry Samuelson.
\newblock Buying locally.
\newblock {\em International Economic Review}, 57(4):1179--1200, 2016.

\bibitem{Parrilo2004SumOS}
Pablo~A. Parrilo.
\newblock Sum of squares programs and polynomial inequalities.
\newblock 2004.

\bibitem{rosenblatt1961principles}
Frank Rosenblatt.
\newblock Principles of neurodynamics. perceptrons and the theory of brain
  mechanisms.
\newblock Technical report, Cornell Aeronautical Lab Inc Buffalo NY, 1961.

\bibitem{stiglitz2017overselling}
Joseph~E Stiglitz.
\newblock The overselling of globalization.
\newblock {\em Business Economics}, 52:129--137, 2017.

\bibitem{winfree2017welfare}
Jason Winfree and Philip Watson.
\newblock The welfare economics of ``buy local".
\newblock {\em American Journal of Agricultural Economics}, 99(4):971--987,
  2017.

\bibitem{young2022should}
Carson Young.
\newblock Should you buy local?
\newblock {\em Journal of Business Ethics}, 176(2):265--281, 2022.

\end{thebibliography}
\bibliographystyle{IEEEtran}

\end{document}